# Controlling the Suhl instability: a numerical study

K. Rivkin[1], V. Chandrasekhar[1] and J. B. Ketterson[1,2,3]

1. Department of Physics and Astronomy, Northwestern University

Evanston IL, 60201

2. Department of Electrical and Computer Engineering, Northwestern University

Evanston IL, 60201

3. Materials Research Center, Northwestern University

Evanston IL, 60201

**Abstract**

Magnetization reversal (switching) using either r.f. fields or brute-force precessional switching is currently thought to ultimately be limited by the non-linear excitation of non-uniform spin waves, the so-called Suhl instability. Here we show (numerically, for the case of a sphere) that this instability can be suppressed by choosing the applied field and/or sphere diameter in such a way that the frequencies of the modes that can be excited through non-linear processes are off-resonance. While the results cannot be explained by a traditional model based on plane waves, they can be understood by projecting the actual state onto the small amplitude spin resonant eigenfunctions.

The Suhl instability[1] in driven spin systems is a well known phenomenon. Historically it has been studied for a magnetic system subjected to a continuous r.f. $H_1$ field at a frequency chosen to excite the uniform mode (the mode for which all the spins precesse in phase with each other) where it is found that above some threshold field non-uniform modes are generated. Initially, a parametric pumping process was studied in which modes with half the pumping frequency are generated. Later, processes involving the generation of second and higher harmonics were studied. The parametric pumping process is now referred to as the "first order Suhl instability", while second harmonic generation is called the "second order Suhl instability"; here we will primarily focus on the second of these instabilities, in which case we will refer to it simply as the Suhl instability.

The Suhl instabilities are currently acquiring a new level of significance due to i) interest in magnetization reversal using time-dependent fields[2,3], ii) a.c. spin polarized currents[4] and iii) so-called "precessional switching"[5] – switching performed by applying an external dc field perpendicular to the direction of magnetization. The effectiveness of a uniform rotation of magnetization, which is the basis for all such processes, can be greatly decreased if non-uniform spin modes are excited in the system[6]. In fact the excitation of non-uniform spin waves is thought to ultimately limit switching behavior in magnetic media[7,8,9]. We note that similar effects, such as spin turbulence, have also been discovered in NMR experiments[10].

In this article we will present numerical studies of various instabilities for the case of a magnetic sphere in a strong external d.c. magnetic field with a smaller r.f. field applied perpendicular to the direction of the d.c. field. We will show that in such a system the Suhl instability can be suppressed by varying the size of the sphere and the value of the applied field. Suppression of this instability may even allow one to use r.f. fields for magnetization reversal. We will further show that the various phenomenon cannot be completely understood using the traditional models of the Suhl instability, but can be explained using numerical methods.

The traditional description of the Suhl instability typically involves a spherical sample but then goes on to assume that the modes are plane waves (whereas even in a uniformly magnetized sphere they are the so-called Walker[11] modes in the absence of exchange). By including non-linear terms in the Landau-Lifshitz[12] equation,

$$\frac{d\mathbf{m}}{dt} = -\gamma \mathbf{m} \times \mathbf{h} - \frac{\beta\gamma}{M_s}\mathbf{m} \times (\mathbf{m} \times \mathbf{h}) \tag{1}$$

where $\gamma$ is a gyromagnetic coefficient, $\mathbf{m}$ is a magnetization, and $\mathbf{h}$ is a magnetic field, one then calculates how the uniform Larmor mode will couple to (pump) modes with frequencies $\omega/2$ (first order Suhl instability) and $2\omega$ (second order Suhl instability). While it is clear from the equations that higher order processes are possible, which create modes with other frequencies, the approach quickly becomes analytically intractable. The model can, with some level of precision, predict the power levels for which the first and second order processes occur; it is also capable of predicting the subsidiary high power resonance (for high r.f. fields one can observe a subsidiary resonance[13] at $\omega/2$) and explaining the lowering of the instability threshold when the sample is simultaneously pumped at $\omega$ and $\omega/2$.[14] However the various assumptions (particularly the plane wave approximation), together with the fact that despite these approximations the equations remain extremely complex, limits the utility of such a mode-mode-coupling based model. We note the Suhl model assumes a continuum of modes within the system (although lying within bands). In a finite system the mode frequencies, and hence their availability to be coupled to, depend on sample size (and the magnetic field).

In this section we describe a numerical study[15] of an Al-substituted YIG sphere using a 4$^{th}$ order Runge-Kutta method to integrate the Landau-Lifshitz equation. We have chosen this material because of its extremely low saturation magnetization, which in turn requires relatively small r.f. fields to observe certain phenomena. However, most of the results and methods obtained here are general and can be applied to other materials.

We assumed the following parameters for the spherical sample: saturation magnetization $M_s$ = 60 Oe, exchange stiffness $A = 5 \times 10^{-7}$ erg/cm, damping constant $\beta = 0.001$; higher values of the damping constant (up to approximately 0.01) lead to very similar results. The d.c. field $\mathbf{H}_0$ is applied along z axis and a circular polarized $\mathbf{H}_1(t)$ r.f. field of the form

$$\begin{aligned} H_{1x}(t) &= H_1 \cos(\omega t) \\ H_{1y}(t) &= H_1 \sin(\omega t) \end{aligned} \quad (2)$$

is assumed, where $H_1$ = 50 Oe; other forms of the r.f. field prevent one from using analytical models.

At the start of each calculation we solve for the equilibrium state of the system. Due to the fact that the external d.c. magnetic field is many times larger than the saturation magnetization, the equilibrium configuration corresponds to the individual moments being almost exactly parallel to z axis. In all cases the frequency of the applied

r.f. field is the Larmor frequency, $\gamma H_0$, which corresponds to the small-amplitude resonance frequency of a sphere.

When the Suhl instability can be neglected (the dipole-dipole interaction is insignificant) and the damping is small enough, the system oscillates back and forth between states corresponding to $M_z / M_s = \pm 1$. However in real systems the onset of the Suhl instability prevents such oscillations; at a certain point the rotation becomes nonuniform. This phenomena creates significant obstacles for utilizing an r.f. field to switch the magnetization direction in magnetic memory devices and for conducting high power FMR experiments.

Various experiments and our calculations have revealed that when sufficiently high r.f. fields are applied the following phenomena appear:

i) The first and second order Suhl instabilities set in, followed by the excitation of higher order modes.

ii) At a certain point, the power starts to transfer back and forth between modes, resulting in auto-oscillations.

iii) After some time, the system becomes chaotic[16,17].

In order to investigate the effectiveness of such processes as a function of the assumed parameters we calculated the minimum value of $M_z / M_s$ that the system can reach for different sphere diameters; if Suhl instability is suppressed, this value should be $-1$. Calculations were carried out for two values of the $H_0$ field: – 600 and 900 Oe. There are two regimes for which the results can be at least qualitatively explained by the use of traditional models:

i) <u>Extremely small spheres (less than 500nm)</u> For $H_0 = 600$ Oe, the system can be well approximated by the single spin model for the first few magnetization reversals. The Suhl instability becomes less and less apparent with decreasing diameter. This behavior can be explained by the fact that all modes, with the exception of a uniform mode, increase in frequency when the diameter is decreased and at some point there are no modes available at $2\omega$ and higher frequencies. Alternatively one can say that when the ratio of the sphere size to the effective exchange length becomes smaller, and exchange interaction keeps the moments aligned.

ii) <u>Large spheres – more than 1000nm.</u> Larger spheres show a prominent Suhl instability. Here there are many modes available at the required frequencies ($2\omega$, etc.), and it is easy to excite them.

Between these two regimes (as presented in figure 1) the minimuim value of $M_z/M_s$ as a function of the diameter displays an oscillatory behavior. For certain diameters the Suhl instability is extremely prominent, while for others it is nearly completely suppressed, at least for the first few magnetization reversals. Here, depending on the size, there may or may-not be a mode available to couple to.

Theory, based on the plane wave assumption of Suhl is capable of a qualitative description of some of the features presented in figure 1, such as a shift toward smaller diameters (as presented, for one of the minima in figure 2) and a reduction in the accessible values of $M_z$ as $H_0$ is increased; at the same time the simple theory is not capable of explaining the oscillatory nature of the graph. Further calculations of the of the oscillatory behavior of the maxima and minima on $H_0$ reveal a complex behavior that lies beyond the plane wave model.

As mentioned above, most studies conclude that the Suhl instability in spheres could be better understood if Walker modes were used in place of plane waves[18]. However exchange has generally been ignored, making it impossible to quantitatively explain the size dependence displayed in figure 1. On the other hand one might simply take a Fourier time transform of magnetization, hoping that this would reveal the individual modes responsible for the instability. This approach is problematic due to extremely rapid onset of the instability (in our case it takes around 4ns for the transition to a "spin turbulent state"[15]. In addition, one has the possibility that multiple modes can be responsible for the second order Suhl instability (as we will see below); the spectrum will be also be complicated by the presence of transient solutions. These facts make such an analysis extremely complicated.

The strategy we introduce here consists of following steps:

a) We calculate the eigenvalues and eigenvectors of the linearized equations of motion of the un-driven system. We then calculate which modes are excited by the external r.f. field.

b) Using the frequencies obtained from step a), we numerically calculate which modes can be excited in the system via the second order Suhl instability.

c) Without using the results from step b, we expand the motion of the system, as obtained from the numerical integration of the full (non-linear) Landau-Lifshitz equation driven by the external field $H_1(t)$, in terms of the eigen modes obtained from step a). Comparing this spectrum with the one obtained in step b) we can determine whether the second order processes are dominant (this can also serve as a verification of the overall strategy).

There are some potential problems with this strategy: Firstly we are using the small amplitude excitations associated with the equilibrium state, assuming that the shift in these modes relative to the true (precessing) system can be neglected (this is surely not true when the system enters the turbulent state). However the frequencies do not enter the calculations associated with step c); while the spectrum calculated via b) can be enhanced through changes of resonant frequencies, numerical experiments confirm that the overall structure of the spectrum remains intact. We also note that our calculations show that the first order Suhl process does not exist for our chosen parameters; the exchange interaction pushes the frequencies of most of the modes *above* the uniform mode frequency, leaving the second order process dominant.

Some of the algorithms we employ here were introduced earlier by us[19,20,21] and will therefore only be discussed briefly. According to the procedure outlined, we must first calculate the small-amplitude resonant modes of the system where we can write the magnetization and magnetic fields as:

$$\begin{aligned}\mathbf{m}(t) &= \mathbf{m}^{(0)} + \mathbf{m}^{(1)}(0)e^{-i\omega t} \\ \mathbf{h}(t) &= \mathbf{h}^{(0)} + \mathbf{h}^{(1)}(0)e^{-i\omega t}\end{aligned} \quad (3)$$

where $\mathbf{m}^{(0)}$ is the static magnetization, $\mathbf{h}^{(0)}$ is the total static magnetic field (internal and external), $\mathbf{m}^{(1)}(t) = \mathbf{m}^{(1)}(0)e^{-i\omega t}$, and $\mathbf{h}^{(1)}(t) = \mathbf{h}^{(1)}(0)e^{-i\omega t}$ are the small time-dependent parts of magnetization and magnetic fields, i.e., the spin waves. Inserting these forms into Eq. (1) and neglecting nonlinear absorption terms like $\beta \mathbf{m}^{(1)} \times \mathbf{h}^{(1)}$ we obtain the following equation:

$$\frac{d\mathbf{m}^{(1)}}{dt} + \gamma \left[ \mathbf{m}^{(0)} \times \mathbf{h}^{(0)} + \mathbf{m}^{(1)} \times \mathbf{h}^{(0)} + \mathbf{m}^{(0)} \times \mathbf{h}^{(1)} + \mathbf{m}^{(0)} \times \mathbf{h}^{(rf)} \right] +$$
$$+ \frac{\beta\gamma}{M_s} \mathbf{m}^{(0)} \times \left( \mathbf{m}^{(1)} \times \mathbf{h}^{(0)} + \mathbf{m}^{(0)} \times \mathbf{h}^{(1)} \right) \approx -\gamma \left[ \mathbf{m}^{(1)} \times \mathbf{h}^{(1)} + \mathbf{m}^{(1)} \times \mathbf{h}^{(rf)} + \mathbf{m}^{(0)} \times \mathbf{h}^{(rf)} \right]$$
(4)

This nonlinear inhomogeneous equation can be rewritten as an integral equation[21]:

$$m_{i\alpha}^{(1)} = V_{i\alpha}^{(k)} \left[ e^{\lambda^{(k)}t} c_k + e^{\lambda^{(k)}t} \int e^{-\lambda^{(k)}t} V_{l\beta}^{(k)*} g_{l\beta}(t) dt \right]$$
$$g_{l\beta}(t) = -\gamma \varepsilon_{\beta\chi\delta} \left[ m_{l\chi}^{(1)} \times h_{l\delta}^{(1)} + m_{l\chi}^{(1)} \times h_{l\delta}^{(rf)} + m_{l\chi}^{(0)} \times h_{l\delta}^{(rf)} \right]$$
(5)

where $\omega^{(k)}$ are the eigenvalues and $V_{i\alpha}^{(k)}$ are the associated eigenvectors of the linear homogeneous eigenvalue problem – which is easily obtained from Eq.(4) by assuming that the right hand side vanishes. Here and in what follows, roman letters refer to individual cells (the system being approximated as a collection of discrete magnetic cells) and greek letters refer to coordinates. We employ the summation convention.

If the system is originally at the equilibrium ($\mathbf{m}^{(0)} \times \mathbf{h}^{(0)} = 0$) non-linear terms $\mathbf{m}^{(1)} \times \mathbf{h}^{(1)}$ and $\mathbf{m}^{(1)} \times \mathbf{h}^{(rf)}$ can be neglected, and assuming the applied r.f. field has sinusoidal time dependency with frequency $\omega$, the Eq.(5) can be solved exactly:

$$m_{l\varphi}^{(1)} = V_{l\varphi k} \left( C_k^{(\omega)} e^{-i\omega t} + C_k^{(-\omega)} e^{i\omega t} \right)$$
$$C_k^{(\omega)} = -i\gamma V_{i\alpha}^{(k)*} \varepsilon_{\alpha\beta\chi} m_{i\beta}^{(0)} \frac{h_{i\chi}^{(rf)}}{\omega^{(k)} - \omega}$$
(6)

In a non-chaotic situation, the mode frequencies $\omega^{(k)}$ have negative imaginary parts. Because of this, we henceforth neglect all terms of the type $e^{-i\omega^{(k)}t}$, because they die out with time (here we will ignore transient solutions). Therefore the solution given by Eq.(6)

corresponds to the steady state excitation of resonant eigenmodes due to the external r.f. field.

After we have identified which modes are excited by the r.f. field (primarily the uniform mode, with the next mode being 0.8% relative of the uniform mode), we can account for a small correction to $m_{l\phi}^{(1)}$ due to nonlinear processes by iterating Eq.(5) [22]. We limit our discussion to second-order Suhl processes, i.e. those which occurs due to the term $\mathbf{m}^{(1)} \times \mathbf{h}^{(rf)}$; we will neglect the term $\mathbf{m}^{(1)} \times \mathbf{h}^{(1)}$ (involving torques arising from the fields associated with the excitations) and other smaller terms. In the first iteration, the $\mathbf{m}^{(1)} \times \mathbf{h}^{(rf)}$ term on the right side of Eq. (5) will produce an additional magnetic moment due to nonlinear excitation of resonant modes $\Delta m_{l\phi}^{(1)}$:

$$\Delta m_{l\phi}^{(1)} = V_{l\phi}^{(k)} C_{k;0}^{\prime(\omega)} e^{-i2\omega t}$$
$$C_{k;0}^{\prime(\omega)} = -i\gamma V_{i\alpha}^{(k)*} \varepsilon_{\alpha\beta\chi} V_{i\beta}^{(0)} \frac{C_0^{(\omega)} h_{i\chi}^{(rf)}}{\left(\omega^{(k)} - 2\omega\right)} \quad . \tag{7}$$

where $C_{k;0}^{\prime(\omega)}$ measures how strongly the mode k can be excited in a second order process arising from a *combination* of the uniform mode (generated by the r.f. field) and the r.f. field itself; the factor $V_{i\alpha}^{(k)*} \varepsilon_{\alpha\beta\chi} V_{i\beta}^{(0)} h_{i\chi}^{(rf)}$ determines the "overlap" of the excited mode $V_{i\alpha}^{(k)}$ with the torque due to the driven mode $V_{i\beta}^{(0)}$ and external r.f. field. In the analysis based on plane waves this overlap is governed by momentum conservation.

It should be mentioned that since the r.f. field has only x and y components, at least some of the $V_{i\alpha}^{(k)*}$ should have non-zero z components. Despite the fact that the system is originally nearly uniformly magnetized along z direction, and therefore the z components of eigenvectors are extremely small, this model is sufficient to make such excitations possible.

We now examine the two cases treated in the previous section for which D = 775.9 nm and D = 905 nm (corresponding respectively to a maxima and a minima in Figure 1) with $H_0 = 600\,\text{Oe}$. Numerical calculations show that in the first case the second order instability will excite modes with $\omega/2\pi = 3.33\,\text{GHz}$, $\omega/2\pi = 3.36\,\text{GHz}$, $\omega/2\pi = 1.84\,\text{GHz}$, $\omega/2\pi = 2.36\,\text{GHz}$, and so on. While most of the eigenmodes excited in the second case have similar spatial distribution, their frequencies are in general substantially lower, and because of this their excitation strength is extremely small – about 10 times less than in the first case. Some of these conclusions differ from the traditional Suhl model: more than one mode can be excited through a second order process, therefore the optimal strategy to suppress the Suhl instability is to place resonant $2\omega$-frequencies *between* the eigenfrequencies of modes that can be excited by second order Suhl instabilities; this explains the oscillatory behavior seen in Figure 1.

The question now is whether such modes are actually excited in the system and to explore it we will use Runge-Kutta integration of the full equations to calculate the magnetization in the sphere as a function of time. For each time step, we can expand the magnetization in terms of the eigenmodes of the linearized system:

$$\begin{aligned}
m_{i\alpha}^{(1)} &= m_{i\alpha}(t) - m_{i\alpha}(0), \\
m_{i\alpha}^{(1)} &\approx a_k(t) V_{i\alpha}^{(k)} \Rightarrow m_{i\alpha}^{(1)} V_{i\alpha}^{(k)*} = a_k(t).
\end{aligned} \quad (3.11)$$

The absolute value of $a_k(t)$ is clearly a measure how strongly the $k^{\text{th}}$ mode is excited. Some results from this procedure are presented on figure 2 (D = 775.9 nm); only the most strongly excited, out of some 30 to 100 excited modes, are shown here. Note all the modes with the exception of the uniform mode, are those identified earlier as candidates for the second order Suhl instability; however some of the expected modes ($\omega/2\pi = 3.36\,\text{GHz}$) are actually absent. A possible explanation is a frequency shift of the missing modes resulting from the large amplitude precession. One must also keep in mind that at about 4ns the system experiences a transition to the spin turbulent[15] state, at which point we expect our assumptions break down.

We reach the following conclusions:

1. We have confirmed that while a model based on representing spin waves as plane waves can be used to explain certain properties of the Suhl instability, in most cases the phenomena are far too complex for such a simplified model to produce accurate quantitative predictions – for example it is not capable of satisfactorily predicting the size dependence.
2. Unlike a traditional plane wave model, numerical analysis predicts the excitation of many modes by the second order instability alone.
3. By varying the sphere diameter and the applied magnetic field one can suppress the excitation of modes that are responsible for the Suhl and other instabilities. This occurs when a resonant frequency lies safely between the frequencies of the excitable modes.

Our technique is expected to be a powerful tool in the study of high power FMR, magnetic spin-echo experiments, and magnetization reversal.

**Acknowledgments.**

This work was supported by the National Science Foundation under grant ESC-02-24210 and DMR 0244711.

.

# Figure captions:

**Figure 1. Minimal value of $M_z$ as a function of the sphere diameter, D.**

**Figure 2. Diameter for which the minimal value of $M_z$ is a minimum as a function of the dc field.**

**Figure 3. Spectral characteristic $a_k$ as a function of time; D = 775.9 nm.**

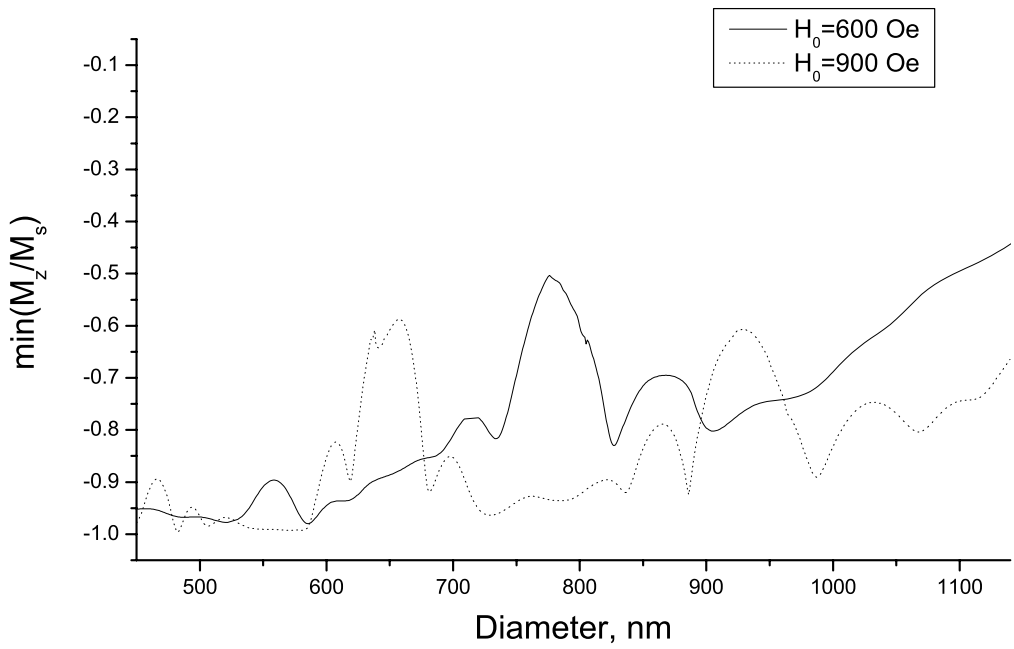

**Figure 1. Minimal value of $M_z$ as a function of the sphere diameter, D.**

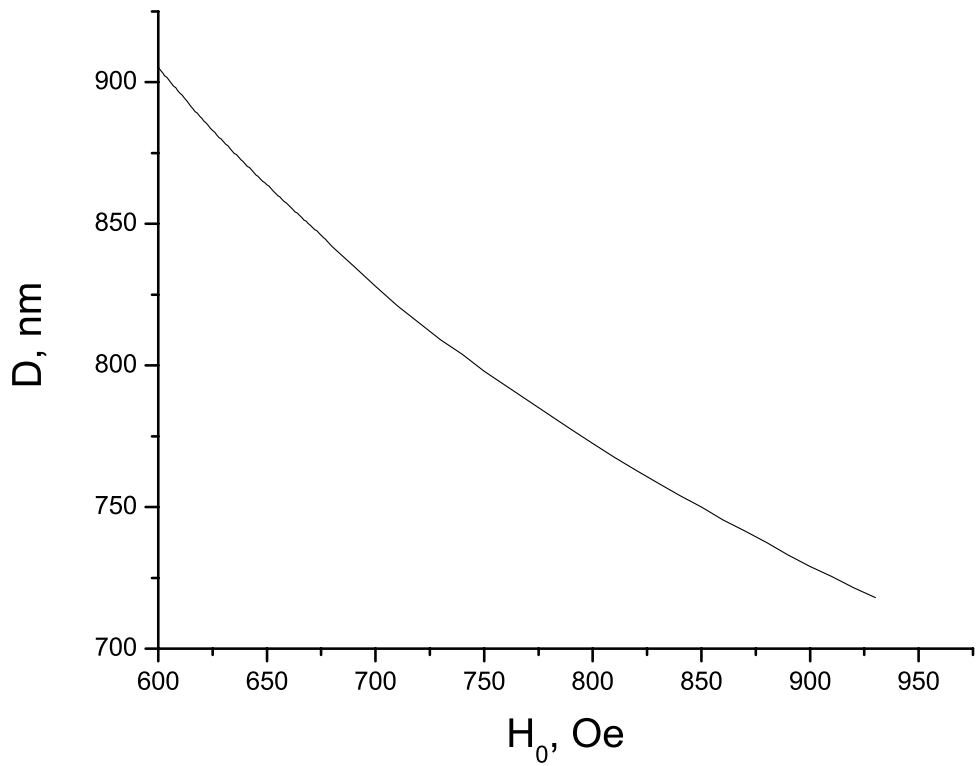

**Figure 2.** Diameter for which the minimal value of $M_z$ is a minimum as a function of the dc field.

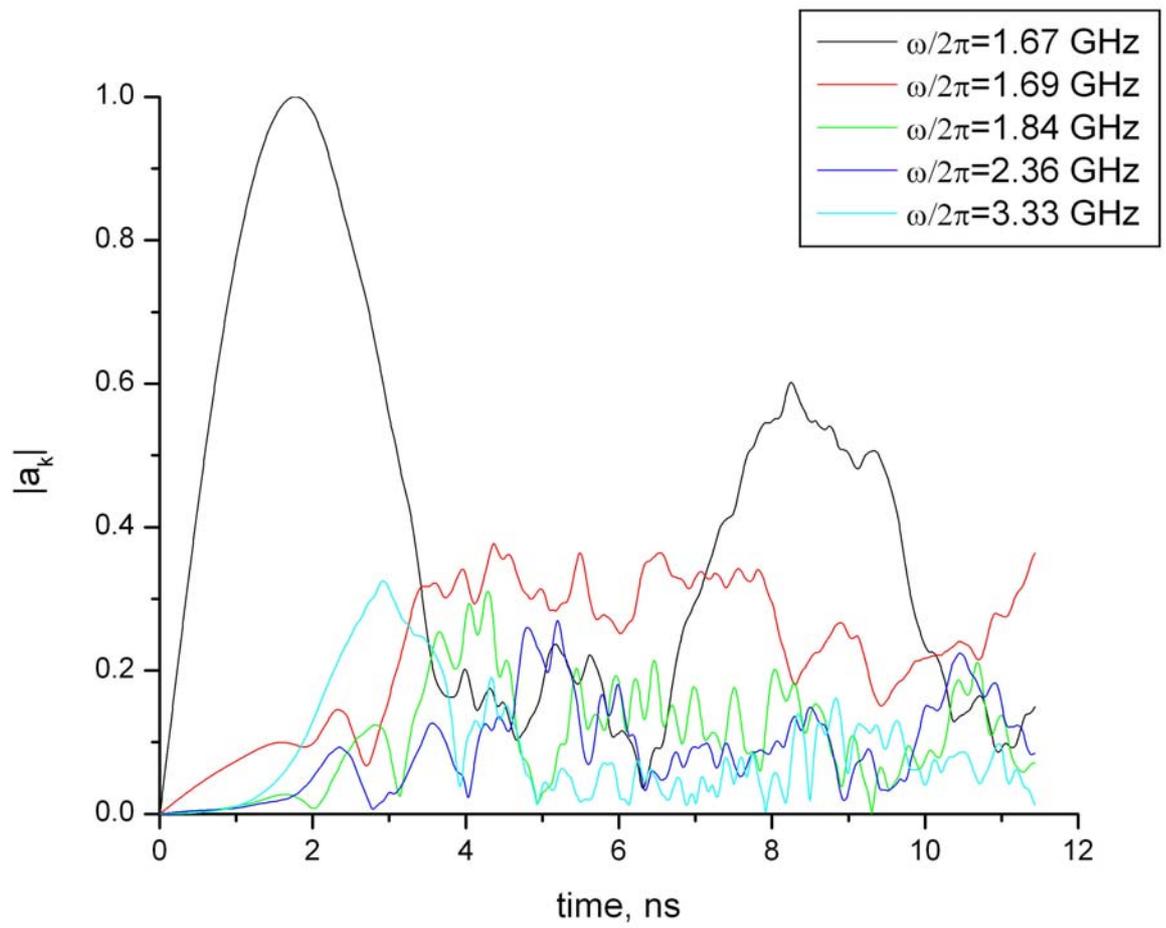

**Figure 3. Spectral characteristic a$_k$ as a function of time; D=775.9 nm.**